\documentclass[english,12pt]{article}
\usepackage[cp1251]{inputenc}
\usepackage{babel}
\usepackage{amsmath, amssymb}
\usepackage{axodraw}
\usepackage{epsfig}
\begin{document}
\title{Contribution of the main radiative corrections to anomalous quartic constants in process
$\gamma\gamma\rightarrow W^+W^-$}
\author{I. Marfin\thanks{BSU, Minsk},\and V. Mossolov\thanks{NC PHEP, Minsk},\and T. Shishkina\footnotemark[1]}
\date{}
\maketitle

\begin{abstract}
Evaluation of  anomalous couplings  in the  $\gamma\gamma\rightarrow W^+W^-$ process   needs to
calculate the cross section $\sigma(W^+W^-)$ with a high precision.
Therefore one has to consider  the main contribution of high order effects.
In this paper contributions of anomalous quartic boson interaction given by
${\cal L}_0$, ${\cal L}_c$, $\tilde{{\cal L}}_0$ are analyzed in high-energy region.
Influence of high order effects is studed in dependences of $\sigma(W^+W^-)$ on anomalous
constants and contour plots with statistical error $2\delta$.
\end{abstract}

\section{Introduction}
$ $

Linear colliders with centre of mass energies running to $1$ TeV
gives a hope for the detection of deviations from SM predictions.
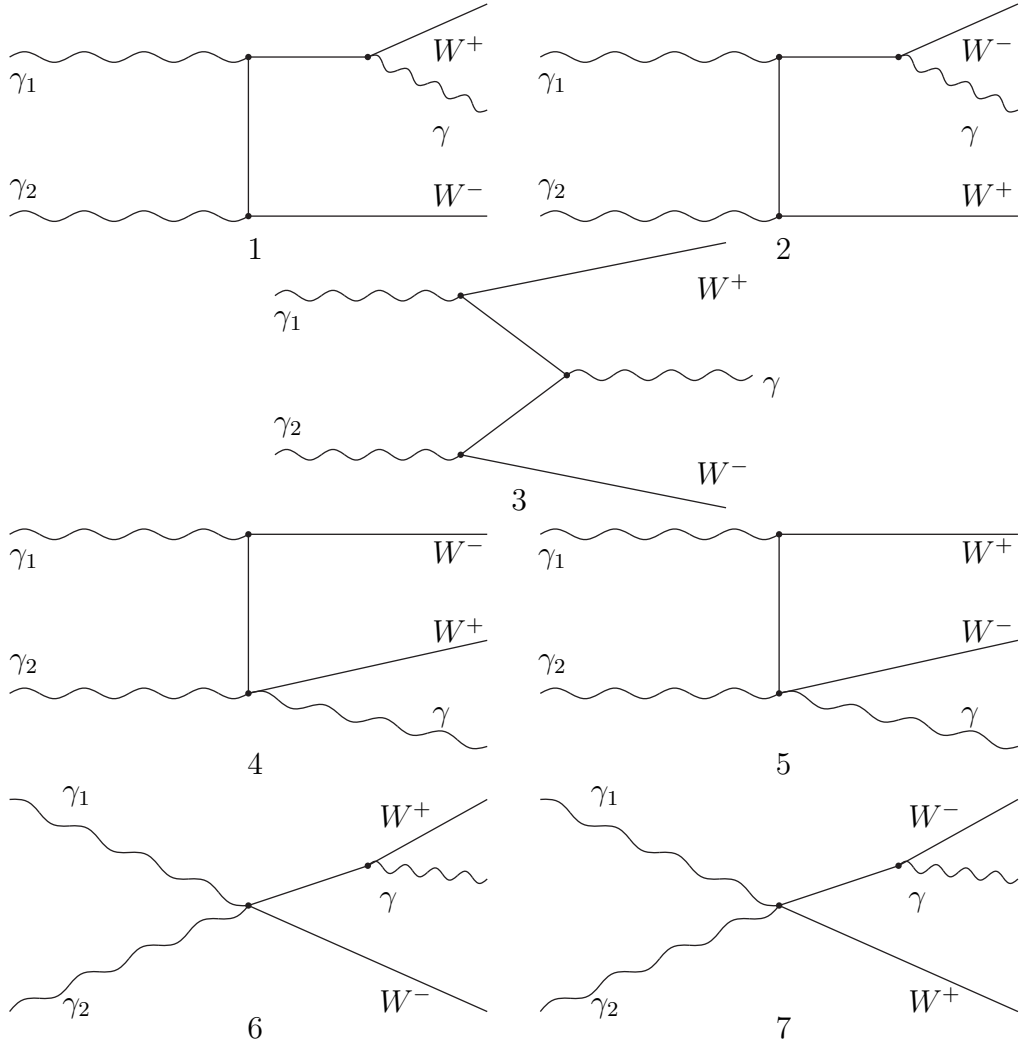
\begin{figure}[h!]
\begin{center}
\begin{picture}(400,400)(0,0)
\Photon(10,370)(100,370){2}{4}\put(10,360){$\gamma_1$}
\Photon(10,310)(100,310){2}{4}\put(10,320){$\gamma_2$}
\Vertex(100,370){1.2}
\Line(100,370)(145,370)
\Vertex(145,370){1.2}
\Line(145,370)(190,390)\put(170,370){$W^+$}
\Photon(145,370)(190,350){2}{4}\put(170,340){$\gamma$}
\Line(100,310)(100,370)
\Vertex(100,310){1.2}
\Line(100,310)(190,310)\put(170,315){$W^-$}
\put(100,295){$1$}

\Photon(210,370)(300,370){2}{4}\put(210,360){$\gamma_1$}
\Photon(210,310)(300,310){2}{4}\put(210,320){$\gamma_2$}
\Vertex(300,370){1.2}
\Line(300,370)(345,370)
\Vertex(345,370){1.2}
\Line(345,370)(390,390)\put(370,370){$W^-$}
\Photon(345,370)(390,350){2}{4}\put(370,340){$\gamma$}
\Line(300,310)(300,370)
\Vertex(300,310){1.2}
\Line(300,310)(390,310)\put(370,315){$W^+$}
\put(300,295){$2$}

\Photon(110,280)(180,280){2}{4}\put(110,270){$\gamma_1$}
\Photon(110,220)(180,220){2}{4}\put(110,230){$\gamma_2$}
\Vertex(180,280){1.2}
\Line(180,280)(280,300)\put(270,280){$W^+$}
\Line(180,280)(220,250)
\Vertex(220,250){1.2}
\Photon(220,250)(290,250){2}{4}\put(295,245){$\gamma$}
\Line(180,220)(220,250)
\Vertex(180,220){1.2}
\Line(180,220)(280,200)\put(270,210){$W^-$}
\put(200,200){$3$}

\Photon(10,190)(100,190){2}{4}\put(10,180){$\gamma_1$}
\Photon(10,130)(100,130){2}{4}\put(10,140){$\gamma_2$}
\Vertex(100,190){1.2}
\Line(100,190)(190,190)\put(170,180){$W^-$}
\Line(100,130)(100,190)
\Vertex(100,130){1.2}
\Line(100,130)(190,150)\put(170,150){$W^+$}
\Photon(100,130)(190,110){2}{4}\put(170,120){$\gamma$}
\put(100,100){$4$}

\Photon(210,190)(300,190){2}{4}\put(210,180){$\gamma_1$}
\Photon(210,130)(300,130){2}{4}\put(210,140){$\gamma_2$}
\Vertex(300,190){1.2}
\Line(300,190)(390,190)\put(370,180){$W^+$}
\Line(300,130)(300,190)
\Vertex(300,130){1.2}
\Line(300,130)(390,150)\put(370,150){$W^-$}
\Photon(300,130)(390,110){2}{4}\put(370,120){$\gamma$}
\put(300,100){$5$}

\Photon(10,90)(100,50){2}{4}\put(30,90){$\gamma_1$}
\Photon(10,10)(100,50){2}{4}\put(30,10){$\gamma_2$}
\Vertex(100,50){1.2}
\Line(100,50)(190,10)\put(150,10){$W^-$}
\Line(100,50)(145,65)
\Vertex(145,65){1.2}
\Line(145,65)(190,90)\put(150,80){$W^+$}
\Photon(145,65)(190,60){2}{4}\put(150,50){$\gamma$}
\put(100,0){$6$}

\Photon(210,90)(300,50){2}{4}\put(230,90){$\gamma_1$}
\Photon(210,10)(300,50){2}{4}\put(230,10){$\gamma_2$}
\Vertex(300,50){1.2}
\Line(300,50)(390,10)\put(350,10){$W^+$}
\Line(300,50)(345,65)
\Vertex(345,65){1.2}
\Line(345,65)(390,90)\put(350,80){$W^-$}
\Photon(345,65)(390,60){2}{4}\put(350,50){$\gamma$}
\put(300,0){$7$}
\end{picture}
\end{center}
\caption{Feynman diagrams of $\gamma\gamma\rightarrow W^+W^-$ accompanied real photon emission }\label{f15}
\end{figure}
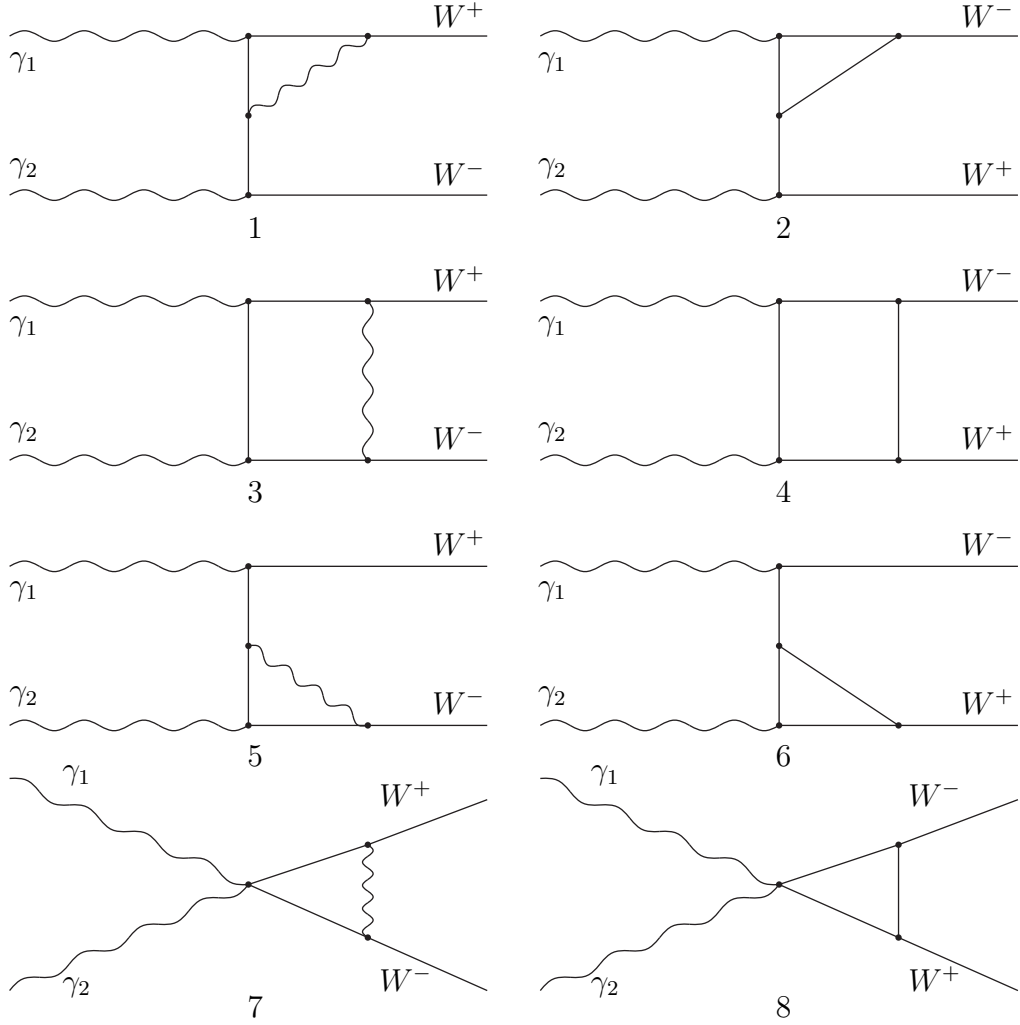
\begin{figure}[h!]
\begin{center}
\begin{picture}(400,400)(0,0)

\Photon(10,370)(100,370){2}{4}\put(10,360){$\gamma_1$}
\Photon(10,310)(100,310){2}{4}\put(10,320){$\gamma_2$}
\Vertex(100,370){1.2}
\Line(100,370)(190,370)\put(170,375){$W^+$}
\Vertex(145,370){1.2}
\Photon(145,370)(100,340){2}{4}
\Vertex(100,340){1.2}
\Line(100,310)(100,370)
\Vertex(100,310){1.2}
\Line(100,310)(190,310)\put(170,315){$W^-$}
\put(100,295){$1$}

\Photon(210,370)(300,370){2}{4}\put(210,360){$\gamma_1$}
\Photon(210,310)(300,310){2}{4}\put(210,320){$\gamma_2$}
\Vertex(300,370){1.2}
\Line(300,370)(390,370)\put(370,375){$W^-$}
\Vertex(345,370){1.2}
\Line(345,370)(300,340)
\Vertex(300,340){1.2}
\Line(300,310)(300,370)
\Vertex(300,310){1.2}
\Line(300,310)(390,310)\put(370,315){$W^+$}
\put(300,295){$2$}
\Photon(10,270)(100,270){2}{4}\put(10,260){$\gamma_1$}
\Photon(10,210)(100,210){2}{4}\put(10,220){$\gamma_2$}
\Vertex(100,270){1.2}
\Line(100,270)(190,270)\put(170,275){$W^+$}
\Vertex(145,270){1.2}
\Photon(145,270)(145,210){2}{4}
\Vertex(145,210){1.2}
\Line(100,210)(100,270)
\Vertex(100,210){1.2}
\Line(100,210)(190,210)\put(170,215){$W^-$}
\put(100,195){$3$}

\Photon(210,270)(300,270){2}{4}\put(210,260){$\gamma_1$}
\Photon(210,210)(300,210){2}{4}\put(210,220){$\gamma_2$}
\Vertex(300,270){1.2}
\Line(300,270)(390,270)\put(370,275){$W^-$}
\Vertex(345,270){1.2}
\Line(345,270)(345,210)
\Vertex(345,210){1.2}
\Line(300,210)(300,270)
\Vertex(300,210){1.2}
\Line(300,210)(390,210)\put(370,215){$W^+$}
\put(300,195){$4$}
\Photon(10,170)(100,170){2}{4}\put(10,160){$\gamma_1$}
\Photon(10,110)(100,110){2}{4}\put(10,120){$\gamma_2$}
\Vertex(100,170){1.2}
\Line(100,170)(190,170)\put(170,175){$W^+$}
\Vertex(145,110){1.2}
\Photon(145,110)(100,140){2}{4}
\Vertex(100,140){1.2}
\Line(100,110)(100,170)
\Vertex(100,110){1.2}
\Line(100,110)(190,110)\put(170,115){$W^-$}
\put(100,95){$5$}
\Photon(210,170)(300,170){2}{4}\put(210,160){$\gamma_1$}
\Photon(210,110)(300,110){2}{4}\put(210,120){$\gamma_2$}
\Vertex(300,170){1.2}
\Line(300,170)(390,170)\put(370,175){$W^-$}
\Vertex(345,110){1.2}
\Line(345,110)(300,140)
\Vertex(300,140){1.2}
\Line(300,110)(300,170)
\Vertex(300,110){1.2}
\Line(300,110)(390,110)\put(370,115){$W^+$}
\put(300,95){$6$}
\Photon(10,90)(100,50){2}{4}\put(30,90){$\gamma_1$}
\Photon(10,10)(100,50){2}{4}\put(30,10){$\gamma_2$}
\Vertex(100,50){1.2}
\Line(100,50)(190,10)\put(150,10){$W^-$}
\Line(100,50)(145,65)
\Vertex(145,65){1.2}
\Photon(145,65)(145,30){2}{4}
\Vertex(145,30){1.2}
\Line(145,65)(190,82)\put(150,80){$W^+$}
\put(100,0){$7$}
\Photon(210,90)(300,50){2}{4}\put(230,90){$\gamma_1$}
\Photon(210,10)(300,50){2}{4}\put(230,10){$\gamma_2$}
\Vertex(300,50){1.2}
\Line(300,50)(390,10)\put(350,10){$W^+$}
\Line(300,50)(345,65)
\Vertex(345,65){1.2}
\Line(345,65)(345,30)
\Vertex(345,30){1.2}
\Line(345,65)(390,82)\put(350,80){$W^-$}
\put(300,0){$8$}
\end{picture}
\end{center}
\caption{Feynman diagrams of one-loop amplitudes of the $\gamma\gamma\rightarrow W^+W^-$}\label{f16}
\end{figure}

The production of two and three electroweak gauge bosons in the
high-energy $\gamma\gamma$ collisions allows to check the anomalous
quartic gauge boson couplings $a_0$, $a_c$, $\tilde{a}_0$ obtained from the Lagrangians
(\ref{l1}) -- (\ref{l3}), for more detailed information see refs. \cite{c4} -- \cite{c3}.
Quartic gauge boson couplings leads to direct  electroweak symmetry breaking,
in  particular to  the scalar  sector of  the theory  or more
generally to new physics of electroweak gauge bosons.
Since the mechanism of symmetry breaking isn't revealed completely so
anomalous quartic gauge bosons can explain it and provide the first evidence of
new physics in this sector of the electroweak theory.
The influence of three possible anomalous couplings on the cross sections of $W^+W^-$
productions has been investigated \cite{c10} at the TESLA kinematics ($\sqrt{S}\sim 1$ TeV).

Correct consideration of results for $\gamma\gamma\rightarrow W^+W^-$
process  and precision analysis of  the future
experiments data are impossible without calculation of whole set the
first-order radiative corrections, presented in figs. \ref{f15},
\ref{f16} (see for example refs. \cite{bib4} -- \cite{bib9}).

Thus at $\sqrt{s} \sim 1\,\, TeV$ the radiative correction
is about a value of Born cross section, it gives contribution to
$\sigma^{Born}(W^+W^-)$ compared with anomalous one.
In this paper we focus our attention  on the most important
QED correction to $\gamma\gamma\to W^+W^-$.

\section{Anomalous Lagrangians of qartic boson interaction}

In order to construct the structures contained anomalous
quartic gauge boson couplings where one photon is involved at
least, one has to consider the operators with the the
lowest dimension of $6$ (see refs. \cite{c2,c1}). That is required for a
custodial $SU(2)_c$
symmetry to have the $\rho=M_W^2/(M_Z^2\cos^2{\theta_W})$
parameter close to $1$. Thus  the $6$-dimensional
operators are considered \cite{c1}
\begin{eqnarray}\label{l1}
\begin{array}{c}
\displaystyle {\cal L}_0 = -\frac{e^2}{16\Lambda^2}a_0F^{\mu\nu}
F_{\mu\nu}\bar{W}^{\alpha}\bar{W}_{\alpha}, \\  \\
\displaystyle {\cal L}_c = -\frac{e^2}{16\Lambda^2}a_cF^{\mu\alpha}
F_{\mu\beta}\bar{W}^{\beta}\bar{W}^{\alpha}, \\  \\
\displaystyle \tilde{\cal L}_0 = -\frac{e^2}{16\Lambda^2}\tilde{a}_0
F^{\mu\alpha}\tilde{F}_{\mu\beta}\bar{W}^{\beta}\bar{W}^{\alpha},
\end{array}
\end{eqnarray}
where we introduce the triplet of gauge bosons
\begin{eqnarray}\label{l2}
\begin{array}{c}
\displaystyle \bar{W}_{\mu} = \left(\frac{1}{\sqrt{2}}(W^+_{\mu}+
W^-_{\mu}),\frac{i}{\sqrt{2}}(W^+_{\mu}-W^-_{\mu}),\frac{1}
{\cos{\theta_W}}Z_{\mu}\right)
\end{array}
\end{eqnarray}
and the field-strenght tensors
\begin{eqnarray}\label{l3}
\begin{array}{c}
\displaystyle F_{\mu\nu} = \partial_{\mu}A_{\nu}
-   \partial_{\nu}A_{\mu}, \\ \\
\displaystyle W_{\mu\nu}^i = \partial_{\mu}W^i_{\nu}
-   \partial_{\nu}W^i_{\mu}, \\ \\
\displaystyle \tilde{F}_{\mu\nu} = \frac{1}{2}
\epsilon_{\mu\nu\rho\sigma}F^{\rho\sigma}.
\end{array}
\end{eqnarray}
The scale $\Lambda$ is introduced to keep the coupling constant
$a_i$ dimensionless \cite{cc3}. In practice,
the $\Lambda$ are specified in the frame of the chosen model for
new physics that supports anomalous quartic
gauge boson couplings. In our case $\Lambda$ are fixed by value
of $M_W$ ($\sim 80$ GeV).
As one can see the operators ${\cal L}_0$ and ${\cal L}_c$
are $C$-, $P$-, $CP$-invariant. $\tilde{{\cal L}}_0$ is the $P$- and $CP$-violating operator.

\section{${\cal O}(\alpha)$ corrections to anomalous constants}

Cross section of the $WW$ pair production in the $\gamma\gamma$ collisions to third order in $\alpha$ is
given by the sum of Born cross section, interference term between of Born and one-loop amplitudes, cross section of
the $WW\gamma$ production. We also include in consideration the process $\gamma\gamma\rightarrow WWZ$  if energies exceed  threshold of $WWZ$ production:
\begin{eqnarray}\label{c15}
\begin{array}{c}
\displaystyle
d\sigma(\gamma\gamma\rightarrow W^+W^-) = d\sigma^{Born}(\gamma\gamma\rightarrow W^+W^-) +
\frac{1}{S}Re(M^{Born}M^{1-loop *})d\Gamma^{(2)} + \vspace{2mm} \\
\displaystyle
+ d\sigma^{soft}(\gamma\gamma\rightarrow W^+W^-\gamma) + d\sigma^{hard}(\gamma\gamma\rightarrow W^+W^-\gamma) +
d\sigma^{Z}(\gamma\gamma\rightarrow W^+W^-Z).
\end{array}
\end{eqnarray}
Real photon emission and one-loop correction are presented in figs. \ref{f15}, \ref{f16}.

Since one-loop and soft photon emission amplitudes are IR-divergent and only their sum is IR-finite it is
convinient to consider soft and hard photon emissions separately \cite{cc1}. $d\sigma^{soft}$ can be presented by factorizable
expression \cite{cc1}
\begin{eqnarray}\label{c16}
\begin{array}{c}
\displaystyle
d\sigma^{soft}(\gamma\gamma\rightarrow W^+W^-\gamma) = d\sigma^{Born}(\gamma\gamma\rightarrow W^+W^-)R^{soft},
\end{array}
\end{eqnarray}
where
\begin{eqnarray}\label{c17}
\begin{array}{c}
\displaystyle
R^{soft} = \frac{2\alpha}{\pi}\left[\left(-1 + \frac{1}{\beta}\left(1-\frac{2M^2_W}{S}\right)\ln\left(\frac{1+\beta}{1-\beta}\right)\right)\left(\ln(2\omega)+ \right.\right. \vspace{2mm} \\
\displaystyle\left. +\frac{1}{n-4} - \ln(2\sqrt{\pi}) + \frac{1}{2}C\right)  + \frac{1}{2\beta}\ln\left(\frac{1+\beta}{1-\beta}\right) + \vspace{2mm} \\
\displaystyle \left. +\frac{1}{2\beta}\left(1-\frac{2M^2_W}{S}\right)\left(Spence\left(\frac{-2\beta}{1-\beta}\right) - Spence\left(\frac{2\beta}{1+\beta}\right)\right)\right],
\end{array}
\end{eqnarray}
$\omega$ is soft photon energy cutoff, $\beta = \sqrt{1-4m_W^2/S}$, $Spence$ -- Spence function and $M_W$ is the mass of $W$ boson.
The differential cross section of hard photon emission is given by
equation:
\begin{eqnarray}\label{c18}
\begin{array}{c}
\displaystyle
d\sigma^{hard}(\gamma\gamma\rightarrow W^+W^-\gamma) = d\sigma(\gamma\gamma\rightarrow W^+W^-\gamma) - d\sigma^{soft}(\gamma\gamma\rightarrow W^+W^-\gamma).
\end{array}
\end{eqnarray}
One-loop amplitude $M^{1-loop}$ can be built due to usage of SCA
(Algebra of Symbolic Calculations) programs ($MATHEMATICA,\,
REDUCE...$) and transformations of scalar and tensor integrals to
one-, two-, ..., six-point integrals. Cross section of $WWZ$
production on $\gamma\gamma$ beams are obtained through aplication
of the Monte-Carlo method of numerical integration and exact
covariant expressions of $\gamma\gamma\rightarrow W^+W^-Z$
amplitudes.
Cross section as well as angular distributions of $WWZ$ production at $\gamma\gamma$ scattering
are presented in our papers \cite{cc1}, \cite{cc2}.

Cross section of the process to third order in
$\alpha$ can be written in the following form
\begin{eqnarray}\label{c19}
\begin{array}{c}
\displaystyle
\sigma(\gamma\gamma\rightarrow W^+W^-) = {\sigma}^{B}+R^{soft}\sigma^{B}+\sigma^{hard} + \sigma^{Z}.
\end{array}
\end{eqnarray}

\begin{figure}[h!]
\begin{minipage}[b]{.975\linewidth}
\centering
\includegraphics[width=\linewidth, height=3.8in, angle=0]{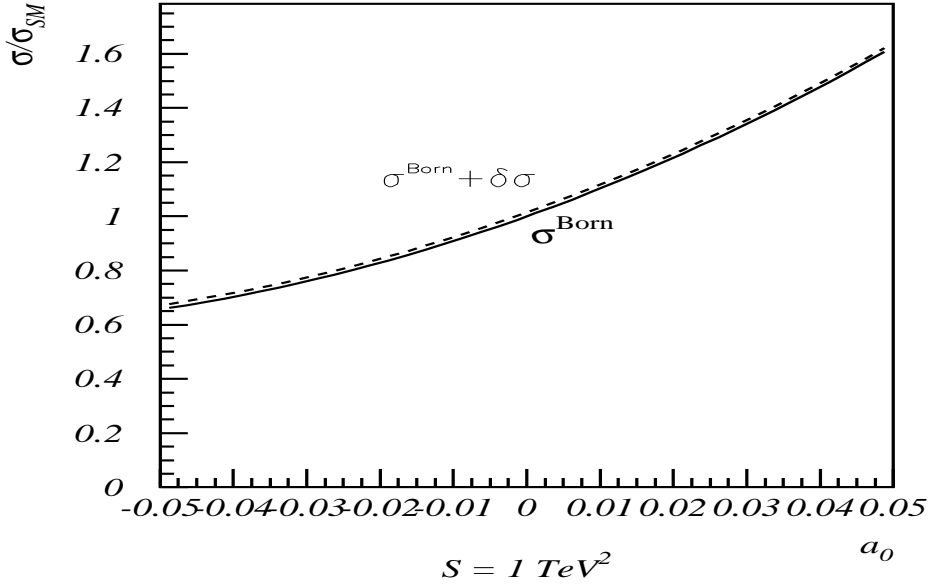}
\caption{Dependence of the cross section
$\sigma(W^+W^-)$ on $a_0$. Solid line presents $\sigma^{Born}$, dashed line -- cross section including radiative correction } \label{f98}
\end{minipage} \hfill
\end{figure}
\begin{figure}[h!]
\begin{minipage}[b]{.975\linewidth}
\centering
\includegraphics[width=1.1\linewidth, height=3.8in, angle=0]{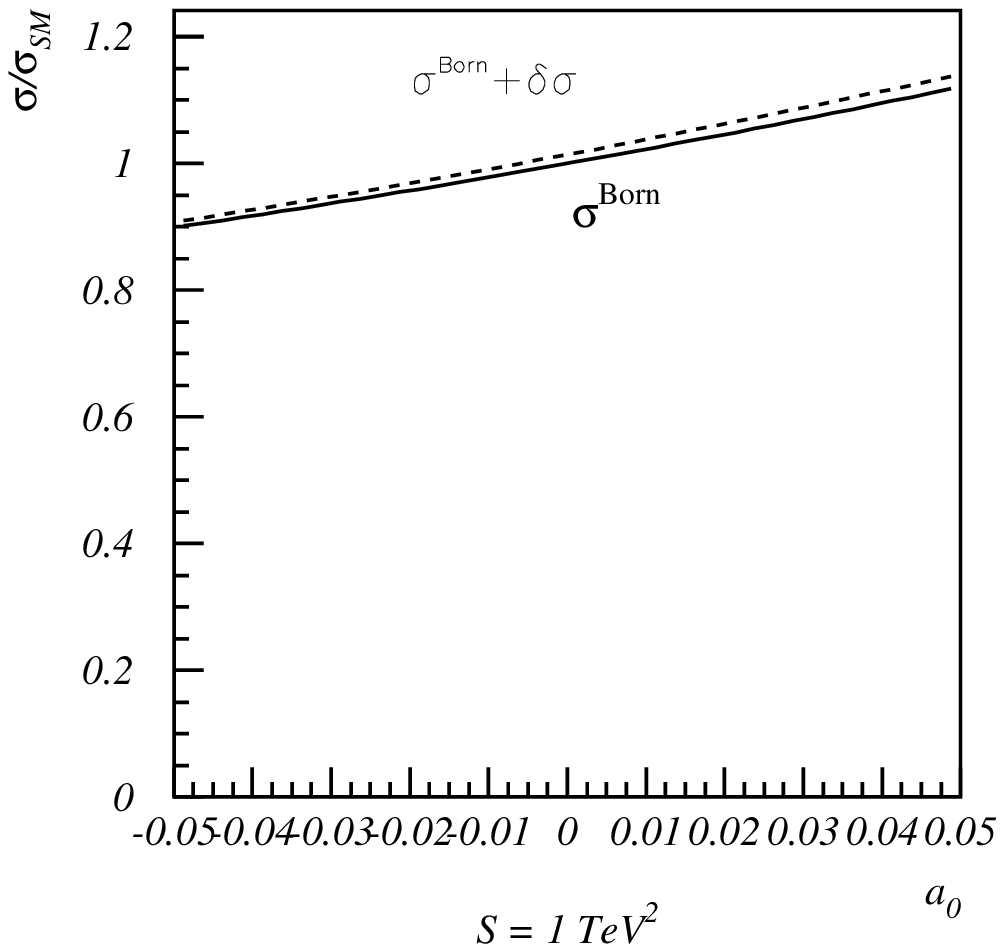}
\caption{Dependence of the cross section
$\sigma(W^+W^-)$ on $a_c$. Solid line presents $\sigma^{Born}$, dashed line -- cross section including radiative correction } \label{f99}
\end{minipage} \hfill
\end{figure}

\begin{figure}[h!]
\begin{minipage}[b]{.475\linewidth}
\centering
\includegraphics[width=1.1\linewidth, height=4.8in, angle=0]{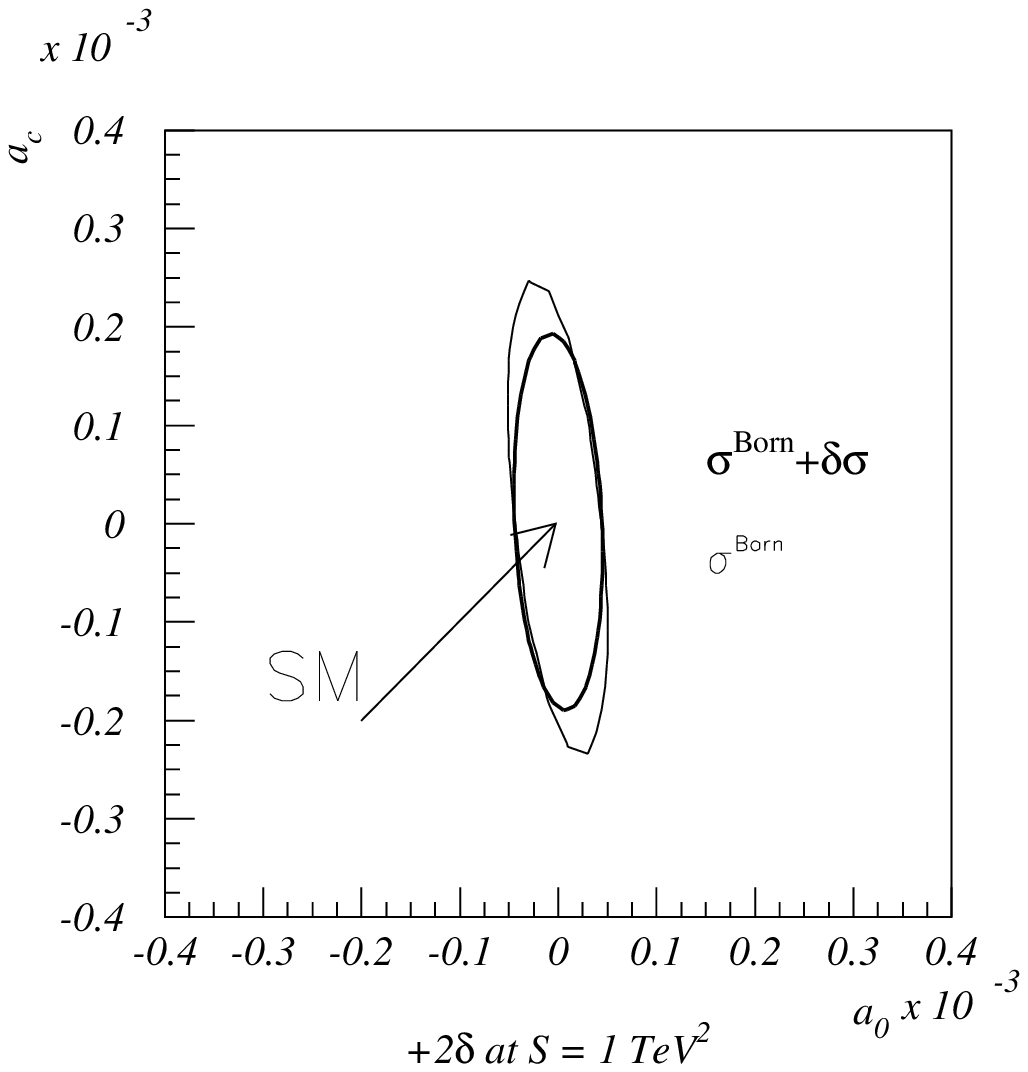}
\caption{ Contour plots on ($a_0$,$a_c$) for $+2\delta$
deviations of $\sigma(W^+W^-)$} \label{f100}
\end{minipage}
\begin{minipage}[b]{.475\linewidth}
\centering
\includegraphics[width=\linewidth, height=4.8in, angle=0]{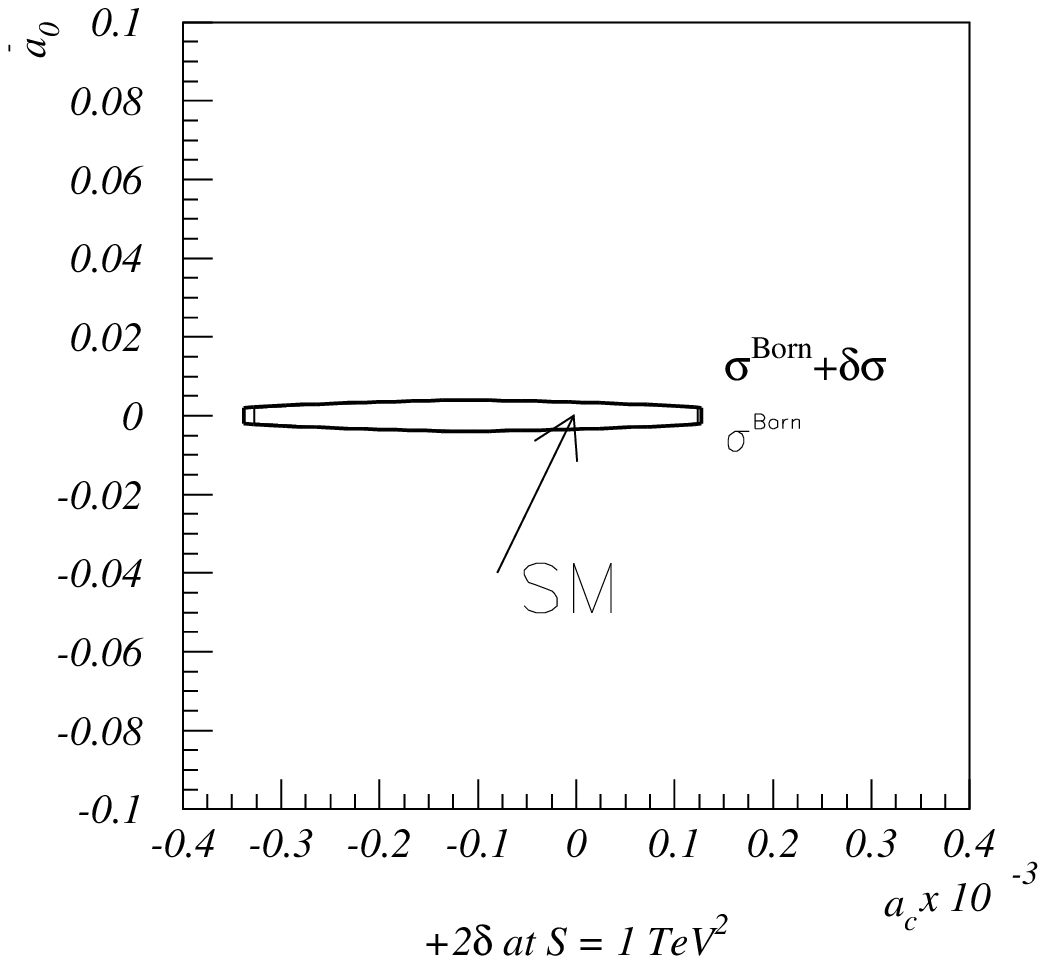}
\caption{ Contour plots on ($a_c$,$\tilde{a}_0$) for $+2\delta$
deviations of $\sigma(W^+W^-)$} \label{f101}
\end{minipage}
\end{figure}

\begin{figure}[h!]
\includegraphics[width=\linewidth, height=5.8in, angle=0]{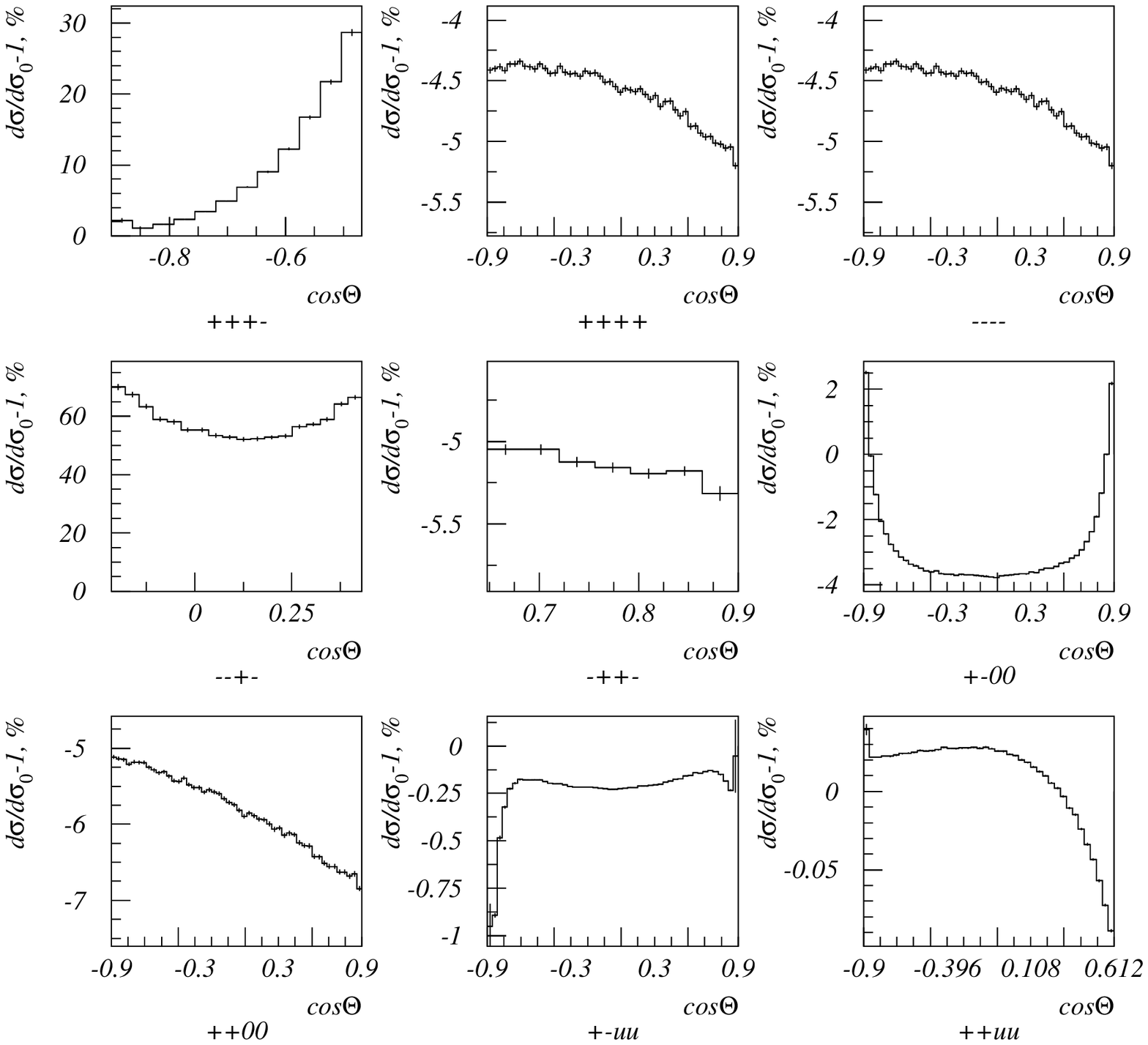}
\caption{Angular distributions of $\delta\sigma$ in $SM+{\cal L}_c$ model,
where +(-) denotes right(left) circular polarization, 0 means longitudinal polarization and
u -- unpolarized case.} \label{f}
\end{figure}

Dependences of total cross section $\sigma(WW)$ on anomalous constant $a_0$ and $a_c$ with radiative correction are presented in figs. \ref{f98}, \ref{f99}.
We see that correction gives appreciable contributions to $\sigma(WW)$ (about $\sim$ $10\%$) only in the case of $a_c$ when the last one is rushing to
the value of $0.05$. For $a_0$ and $\tilde{a}_0$ changes in anomalous part of $\sigma(WW)$ concerned with correction
are negligible. So we conclude that in the first order of perturbation theory it has sense to include at considering of the
anomalous Lagrangian ${\cal L}_c$ only. This fact is confirmed by contour plots presented in the paper (see figs. \ref{f100},\ref{f101}).

In order to evaluate the region of scattering angule where the radiative correction gives a considerable contribution
we need to built graphs of angular distributions for $\delta\sigma$ in the $SM+{\cal L}_c$ model. Figure \ref{f}
shows this dependence.

Taking  into account the value of luminosity ${\cal L}$ of photons
about $100fb^{-1}/year$, energy $\sqrt{S}\sim 1\, TeV$ that
corresponds to TESLA experimental conditions \cite{c10}, we can
build contour plots on $a_0$, $a_c$, $\tilde{a}_0$ of $\sigma(WW)$
with the lowest order radiative correction (see figs. \ref{f100},
\ref{f101}), where  statistical error $\delta$ is equal to
$0.05\%$. Calculation of $\delta\sigma$ leads to fact that the
ellipse built on $(a_0,a_c)$ pair with $+2\delta$ error is
shrinked by factor of $1/4$. This implies that inclusion of
radiative correction in consideration increases confidence level
what improves chances to discover deviations from SM at future
experiments.

\end{document}